\documentclass[%
aip,
cp,  
amsmath,amssymb, 
reprint, %
]{revtex4-2}

\def\Xint#1{\mathchoice
   {\XXint\displaystyle\textstyle{#1}}%
   {\XXint\textstyle\scriptstyle{#1}}%
   {\XXint\scriptstyle\scriptscriptstyle{#1}}%
   {\XXint\scriptscriptstyle\scriptscriptstyle{#1}}%
   \!\int}
\def\XXint#1#2#3{{\setbox0=\hbox{$#1{#2#3}{\int}$}
     \vcenter{\hbox{$#2#3$}}\kern-.5\wd0}}

\def\dashint{\Xint-}

\usepackage{graphicx}
\usepackage{dcolumn}
\usepackage{bm}
\usepackage{floatrow}
\usepackage[utf8]{inputenc}
\usepackage[T1]{fontenc}
\usepackage{mathptmx} 

\begin{document}

\title{$\pi\pi$ scattering from a similarity renormalization group perspective}

\author{Mar\'ia G\'omez-Rocha} 
 \email[Corresponding author: ]{mgomezrocha@ugr.es}
\author{Enrique Ruiz Arriola}%
\email{earriola@ugr.es}
\affiliation{
  Departamento de F\'isica At\'omica, Molecular y Nuclear and Instituto Carlos I de F\'isica Te\'orica y Computacional, Universidad de Granada E-18071 Granada, Spain.
}

\date{\today} 

\begin{abstract}
A Wilsonian approach based on the Similarity Renormalization Group to
$\pi\pi$ scattering is analyzed in the $JI=$00, 11 and 02 channels in
momentum space up to a maximal CM energy of $\sqrt{s}=1.4$ GeV. We
identify the Hamiltonian by means of the 3D reduction of the
Bethe-Salpeter equation in the Kadyschevsky scheme.  We propose a new
method to integrate the SRG equations based in the Crank-Nicolson
algorithm with a single step finite difference so that isospectrality
is preserved at any step of the calculations.  We discuss issues on
the high momentum tails present in the fitted interactions hampering
calculations.
\end{abstract}

\maketitle

\section{Introduction: \\ what are natural scales in a physical problem?}

The renormalization group has been a milestone in the discussion of
relevant scales in quantum field theory and in particular in the study
of strong interactions~\cite{Wilson:1973jj}.  The main reason lies not
only on the difficulty of hadronic physics on its own but also in the
fact that hadronic binding and scattering is a highly non-perturbative
phenomenon. There are several approaches which have been proposed in
the past to deal with this issue and in the present contribution we
advance our findings based on a scheme proposed in the 90's by
Wegner~\cite{wegner2001flow} and simultaneously by G\l azek and
Wilson~\cite{Glazek:1993rc}. These approaches have been applied both in
QCD itself to deal with heavy quarks and gluon binding~\cite{Glazek:2017rwe,Serafin:2018aih}, the
running of the strong coupling constant~\cite{Gomez-Rocha:2015esa} as well as in low
energy Nuclear Physics within the context of nuclear
binding~\cite{Bogner:2001gq,Bogner:2003wn} (for a recent review see
e.g.~\cite{Hergert:2015awm}). Our aim here will be to extend these methods
to low energy hadronic physics and we will consider as a starting step
the case of $\pi\pi$ scattering leaving a more detailed study for a
future work. This is the lowest energy interacting process in hadronic
physics which has been studied in much detail and where there is a
wealth of accurate results as well as a long
history~\cite{Colangelo:2000jc,Ananthanarayan:2000ht,Colangelo:2001df,Caprini:2003ta,Pelaez:2004vs,Kaminski:2006yv,Kaminski:2006qe,GarciaMartin:2011cn}.

Most of the studies concerning $\pi\pi$ interactions have been carried out using separable potentials with long high-momentum tails. 
For instance, we consider here the potential used in Ref.~\cite{Mathelitsch:1986ez} to fit $\pi\pi$ scattering phase-shifts. Those potentials have long tails that go up to 10 or even 100 GeV. This is a disturbing fact if one is taking into account a regime in which pions are structureless objects. A recent study in coordinate space~\cite{RuizdeElvira:2018hsv} displays also these long momentum tails. This suggests that, although the potentials fit very precisely the experimental data, they are of little help in order to provide a deeper physical information in the low-energy regime. 
In this work, we investigate in a preliminar fashion the properties of the SRG method that may help in the study of the $\pi\pi$ interaction in the physical region.

\section{The SRG method}

The application of the Similarity Renormalization Group (SRG) is based
on the definition of a Hamiltonian. In general terms, for a given
Hamiltonian, which will be denoted as $H_0$, the SRG equations are
formally written as a double commutator structure, and an initial
condition
\begin{eqnarray}
\frac{d H_s}{ds}= [[G_s,H_s],H_s] \ , 
\label{eq:srg}
\end{eqnarray}
where $H_0$ is the initial condition and $G_s$ is the generator of
the SRG evolution. Within this context, the parameter which has a
physical interpretation is the so-called similarity cut-off, which we
denote by $\lambda \equiv 1/\sqrt{s}$ and has energy
dimension~\footnote{This is at difference with the non-relativistic
  case~\cite{wegner2001flow,Glazek:1993rc} where $\lambda= 1/s^{\frac14}$ and in RGPEP~\cite{Glazek:2017rwe,Serafin:2018aih,Gomez-Rocha:2015esa}
  where $\lambda= 1/s$.}. The simplest choice is to take $G_s=T$ which
corresponds to the G\l azek-Wilson case. One property of the SRG is
that the evolved Hamiltonian $H_s$ has the same spectrum as the
original one. Actually, in the limit $s \to \infty$, the Hamiltonian
becomes diagonal in the basis where the generator $T$ is also
diagonal.  Therefore the SRG method implements a diagonalization of
the Hamiltonian in a continuous fashion rather than in the finite
number of steps which are usually involved in a numerical
diagonalization procedure such as the Gauss elimination method (see e.g. discussions in Refs.~\cite{Arriola:2013gya,Arriola:2016fkr}).

\section{Kadyshevsky equation}

Unlike the more customary case of $NN$ scattering where for many
practical purposes the non-relativistic formalism applies, in the
$\pi\pi$ case the genuinely non-perturbative aspects of the
interaction manifest themselves at energies where relativity becomes
essential. Indeed, the occurrence of resonances such as the $\rho$ and
$\sigma$ mesons, fulfill $m_\rho,m_\sigma \gg 2 m_\pi$, the CM
threshold energy. Although the standard approach in this case would be
the Bethe-Salpeter equation~\cite{Salpeter:1951sz} we prefer to
describe the scattering problem, in terms of the Kadyshevsky
equation~\cite{Kadyshevsky:1967rs}. This is a 3D-reduction of the Bethe-Salpeter
equation that enables a relativistic Hamiltonian interpretation for
the scattering problem.  Furthermore, at the partial-waves level, they
reduce to 1D linear integral equations which can be handled with a
moderate numerical effort. As compared to other 3-D
approaches~\cite{polivanov1964spectral}, this particular 3-D reduction
satisfies a Mandelstam representation, i.e. a double dispersion
relation both in the invariant mass $s$ and momentum $t$ Mandelstam
variables~\cite{Skachkov:1970ia}. The appearance of spurious
singularities has been addressed in the different approaches in
Ref.~\cite{Yaes:1973kj}. In addition, the Kadyshevsky equation also
lacks spurious singularities in the related three-body
problem~\cite{Garcilazo:1984rx}.  Actually, there has been already
some work with this equation for the case of $\pi\pi$
scattering~\cite{Mathelitsch:1986ez} for separable potentials, where
the lowest partial waves corresponding to 
$S$, $P$ and $D$ angular momenta
have been fitted. This will be discussed below in more detail.

The Kadyshevsky equation reads~\cite{Kadyshevsky:1967rs}
\begin{eqnarray}
t(\vec p', \vec p, \sqrt{s}) =  v(\vec p', \vec p)  + \int \frac{d^3 q}{(2\pi)^3}
\frac{v(\vec p',\vec q)}{4 E_q^2} 
\frac{t(\vec q,\vec p, \sqrt{s})}{\sqrt{s}-2 E_q + i \epsilon} \ ,  \\
\end{eqnarray}
where $t(\vec p', \vec p, \sqrt{s})$ is the transition amplitude. 
The potential is symmetric $v(\vec p', \vec p)=v(\vec p, \vec
p')$ and energy {\it independent}. Using rotational invariance, we can write
\begin{eqnarray}
  t(\vec p', \vec p, \sqrt{s}) = 4 \pi \sum_{lm} Y_{lm} (\hat p') Y_{lm}
  (\hat p)^* t_l (p',p, \sqrt{s}) \ ,
\end{eqnarray}
so that, the partial waves level and for spin zero equal mass particles we get
\begin{eqnarray}
  t_l (p',p,\sqrt{s}) &=&  v_l(p',p)   + \int_0^\infty dq \, \frac{q^2}{4 E_q^2} 
  \frac{v_l(p',q)t_l (q,p, \sqrt{s})}{\sqrt{s}-2 E_q + i \epsilon}  \ , 
  \label{Eq:Kad} 
\end{eqnarray}
where $+i \epsilon$ implements the Feynman boundary condition, $E_q =
\sqrt{q^2+m_\pi^2}$ is the intermediate energy and, on the mass shell, one has $\sqrt{s} = 2
\sqrt{p^2+m_\pi^2}$ with $p$ being the center of mass (CM) momentum.

For a real potential
this equation satisfies the two-body unitarity condition, so that the
phase-shift is given by
\begin{eqnarray}
  -\tan \delta_l(p) = \frac{\pi}{8} \frac{p}{E_p} r_l (p,p,\sqrt{s} ) 	\ ,
  \label{eq:phLS}
\end{eqnarray}
where $r_l$ is the corresponding reaction matrix satisfying 
\begin{eqnarray}
r_l (p',p,\sqrt{s}) &=& v_l(p',p) + \dashint_0^\infty dq \, \frac{q^2}{4 E_q^2} 
\frac{v_l(p',q) r_l (q,p, \sqrt{s})}{\sqrt{s}-2 E_q }  \ ,
\end{eqnarray}
and the principal value has been introduced in the integral.

\section{Numerical results for a Separable Model}

The advantage of using a 3D reduction of the BS equation is
the existence of a Hamiltonian interpretation. The Hamiltonian version 
of the Kadyshevsky equation reads
\begin{eqnarray}
H \Psi_l (p) &\equiv& 2 E_p \Psi_l(p) 
+ \int_0^\infty dq \frac{q^2}{ 2E_q^2} v_l (p,q) \Psi_l(q) \ , \nonumber \\
& = & \sqrt{s} \Psi_l(p) \ .
\end{eqnarray}
this equation will be explicitly used below in the SRG formalism. 

\subsection{The model}

For simplicity we use here the separable model 
of Garzilazo and Mathelitsch~\cite{Mathelitsch:1986ez}
\begin{eqnarray}
  V_\alpha (p,p') = \eta_\alpha g_\alpha (p) g_\alpha (p') \ ,
\end{eqnarray}
where the subscript $\alpha=IJ$ indicates the channel, 
and the form factors $g_\alpha(p)$ are given by
\begin{eqnarray} 
  g_{00}(p) &=&\frac{617.865 p^2}{\left(p^2+99.3951\right)^2}+\frac{423.64}{p^2+1034.75}
   \ ,  \\
g_{11}(p)  &=&p \left[\frac{132.237}{p^2+900.462} - \frac{5.11596}{p^2+21.9744}\right]  \ , \\
g_{02}(p) &=& \frac{3.65 p^2}{\left(p^2+3.9601\right)^2} + \frac{175.7}{p^2+357.21} \ ,
\end{eqnarray}
and the signs corresponding to attractive ($\eta < 0$) or repulsive ($\eta > 0$) interactions are 
\begin{eqnarray}
  \eta_{00} & = & -1 \ ,  \quad \eta_{11} \, = \, -1 \ ,  \quad \eta_{02} \, = \, 1 \ .
\end{eqnarray}
The parameters in the $g_\alpha$'s have been refitted to describe the upgraded Madrid analysis~\cite{GarciaMartin:2011cn}.
One important aspect of these separable potentials regards the long tails which extend to unrealistic values of CM momentum 
$p \sim 10-200$GeV which need to be handled with care in the numerical analysis. The analytical solution for this separable model is solved by the ansatz
\begin{eqnarray}
t_l(p',p, \sqrt{s})= g_l(p') g_l(p) t_l (\sqrt{s}) \ ,
\end{eqnarray}
and inserting this in Eq.~(\ref{Eq:Kad}) we get
\begin{eqnarray}
\left[t_l( \sqrt{s}) \right]^{-1} = 1 - \int_0^\infty dq \,
\frac{q^2}{4 E_q^2} \frac{ \eta  \left[g_l(q)\right]^2}{\sqrt{s}-2
  E_q } \ ,
\end{eqnarray}
yielding the final result 
\begin{eqnarray}
p \cot \delta_l(p) = - \frac{8 E_p}{\pi v_l(p,p)} \left[ 1 - \dashint_0^\infty dq \,
\frac{q^2}{4 E_q^2} \frac{v_l(q,q)}{\sqrt{s}-2 
  E_q } \right] \ .
\end{eqnarray}

\begin{figure}
\includegraphics[scale=0.8]{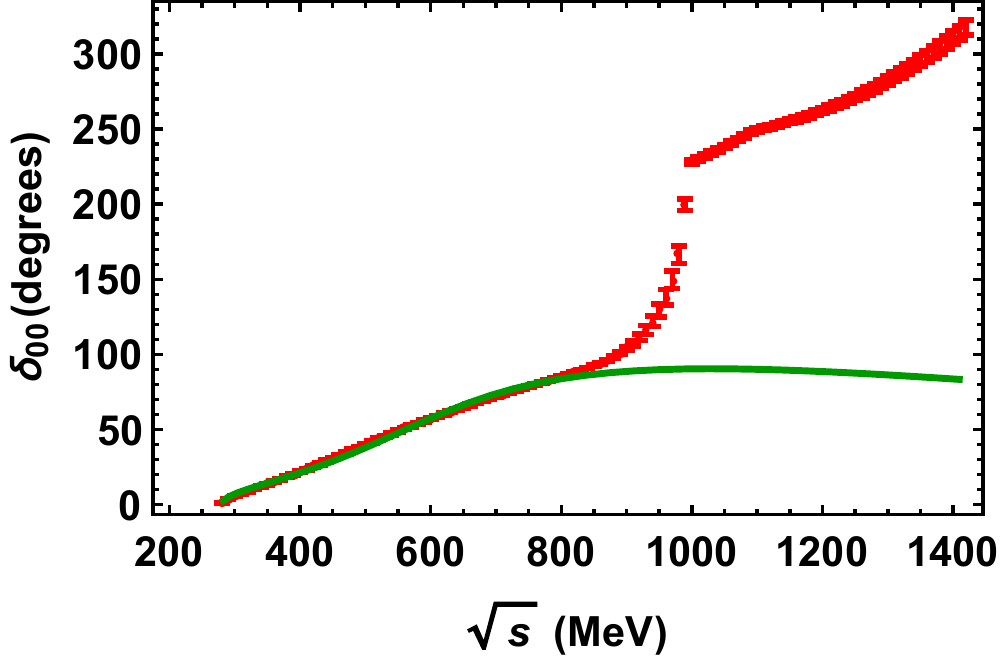}
 \includegraphics[scale=0.87]{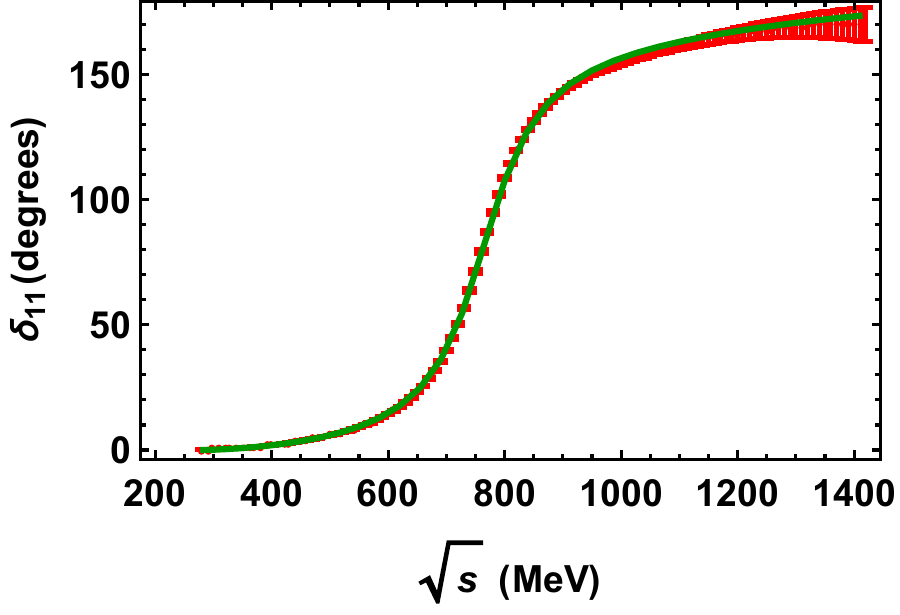}

\floatbox[{\capbeside\thisfloatsetup{capbesideposition={right,center},capbesidewidth=7.5cm}}]{figure}[\FBwidth]
{\caption{Colour online: Phase shifts for $\pi\pi$ scattering for the
  $JI=00,11,02$ channels as a function of the CM energy  comparing
  the analytical separable solution with the energy-shift using a grid
  with $N=100$ points. The data and their uncertainties (in red) are taken from
  Ref.~\cite{GarciaMartin:2011cn}.}\label{Fig:phases}}
{\includegraphics[scale=0.86]{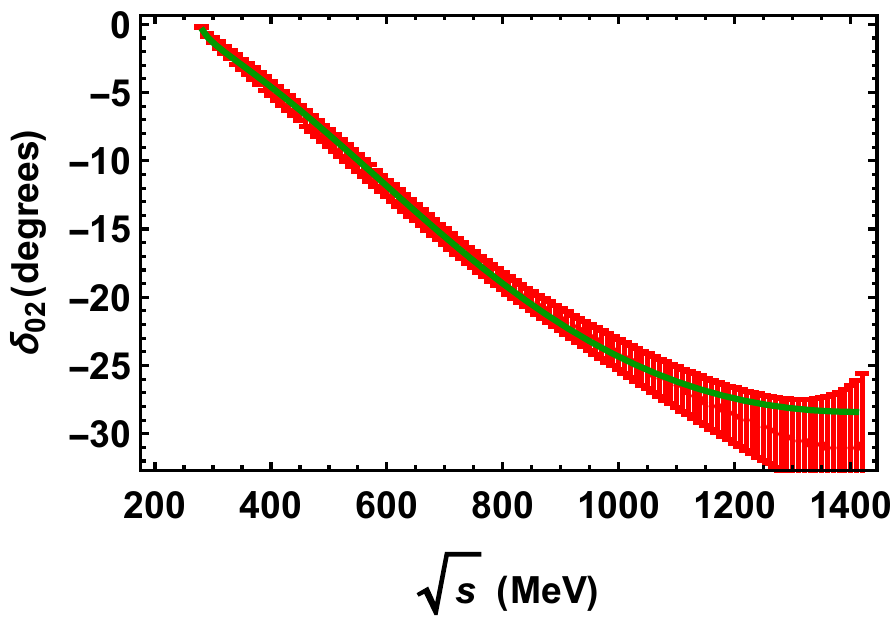}}
\end{figure}

Figure~\ref{Fig:phases} shows the phase shifts using this model and compared to the experimental upgrade of the Madrid group~\cite{GarciaMartin:2011cn} and as we see the fit is rather reasonable, displaying the most conventional features such as the $\sigma$ and $\rho$ resonances in the 00 and 11 channels respectively. As usual, in the 00 channel we see a rising around the 1000 MeV, which correspond to the onset of the $K \bar K$ threshold. Our fit for the 00 channel differs above this energy, since we are not considering this inelastic effect in our potential or the $f_0(960)$ resonance. 

\subsection{SRG evolution}
Once we have fixed our Hamiltonian we can directly proceed to implement the SRG equations. 
If, for definiteness, we focus on the Wilson generator, we get
\begin{eqnarray}
  \frac{d V_s (p',p)}{ds} &=& -(2E_{p`}-2 E_p)^2 V_s (p',p) \nonumber
  + \int_0^\infty dq \frac{q^2}{4 E_q^2} V_s(p',q) V_s(q,p) (2
  E_p + 2 E_{p'}-4 E_q) \ ,
\end{eqnarray}
where we have taken the generator to be the relativistic kinetic energy $\sqrt{p^2 + m_\alpha^2}=E_q$ 
and sandwiched Eq.~(\ref{eq:srg}) between free CM momentum states. 
These are complicated non-linear integro-differential equations which become
numerically messy. At large momenta we can neglect the non-linear term
and hence we get the solution
\begin{eqnarray}
  V_s (p',p) \sim e^{-s (2E_{p`}-2 E_p)^2} V_0 (p',p) \ ,
\end{eqnarray}
which suggests that the effect for SRG evolving is narrowing the interaction to a region of a width $\sim \lambda$.
In order to perform the evolution, we use the Crank-Nicolson algorithm~\cite{crank1947practical,crank1996practical}, in an analogous way as it is used in the time evolution of states that satisfy the Schr\"odinger equation. We will provide more details of the advantages of this method in an upcoming work~\cite{inpreparation}. Note also that the long tails described above for the separable potentials requires a rather large Hilbert space in order to integrate the SRG equations. 

\begin{figure*}
\centering
\includegraphics[scale=0.56]{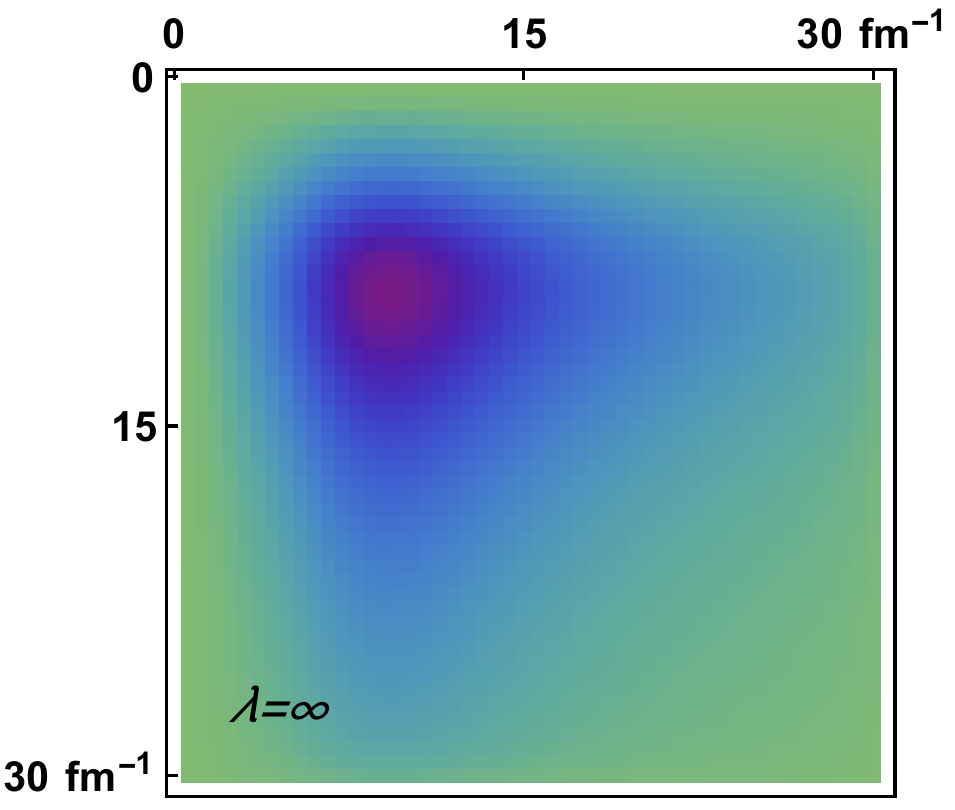}
\includegraphics[scale=0.56]{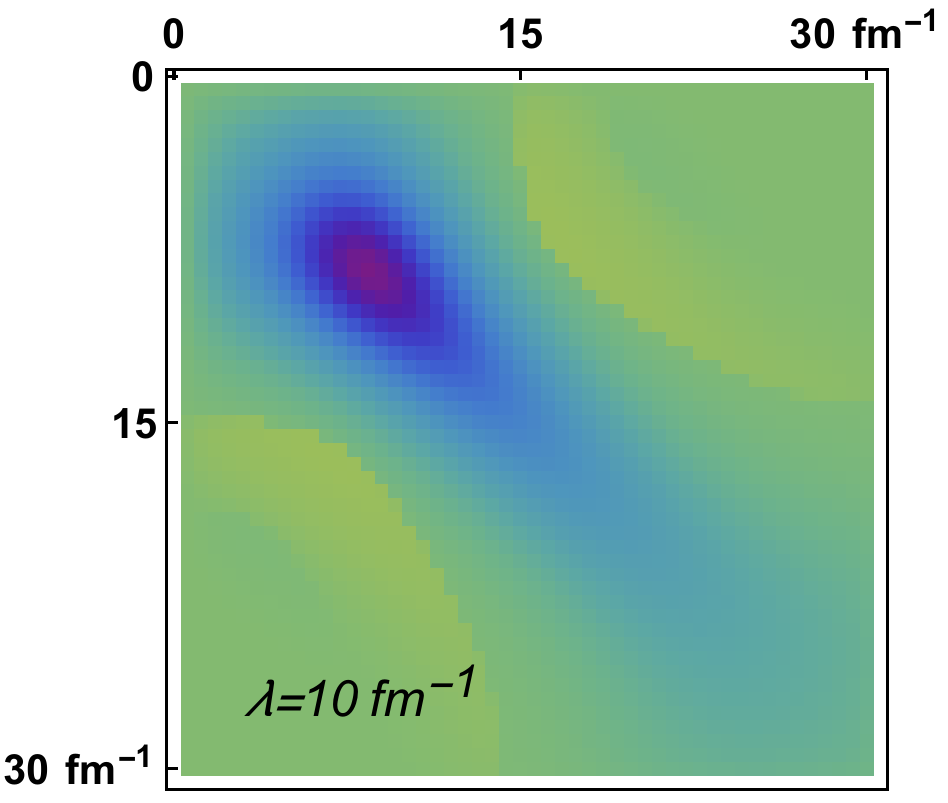}
\includegraphics[scale=0.56]{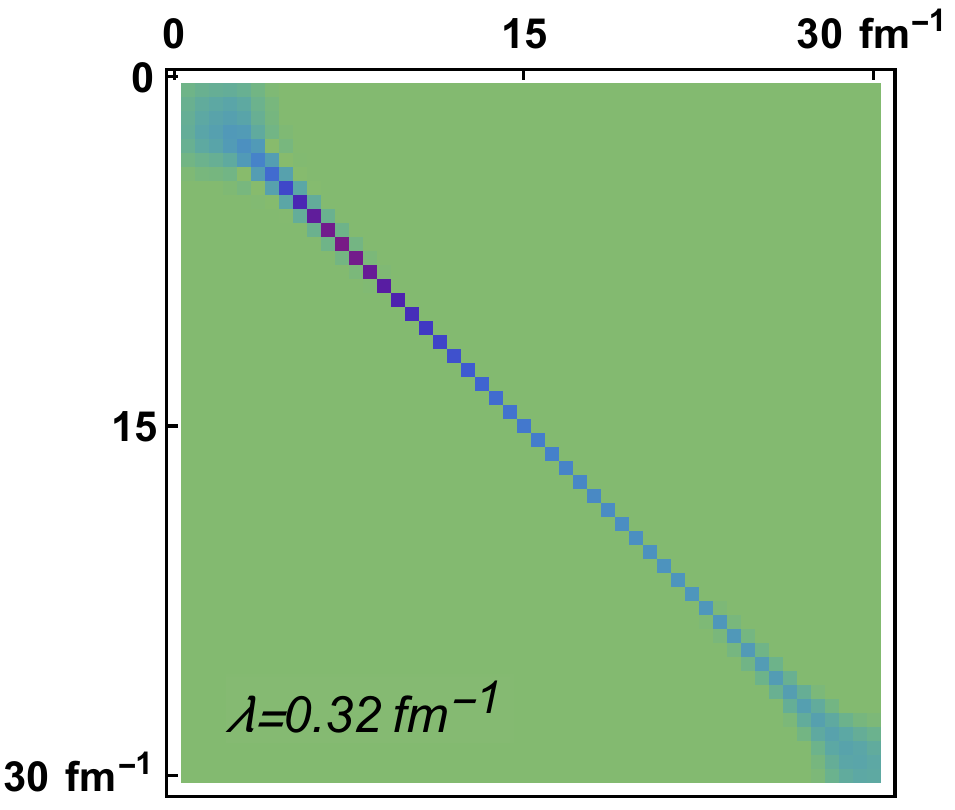} 
\\
\includegraphics[scale=0.55]{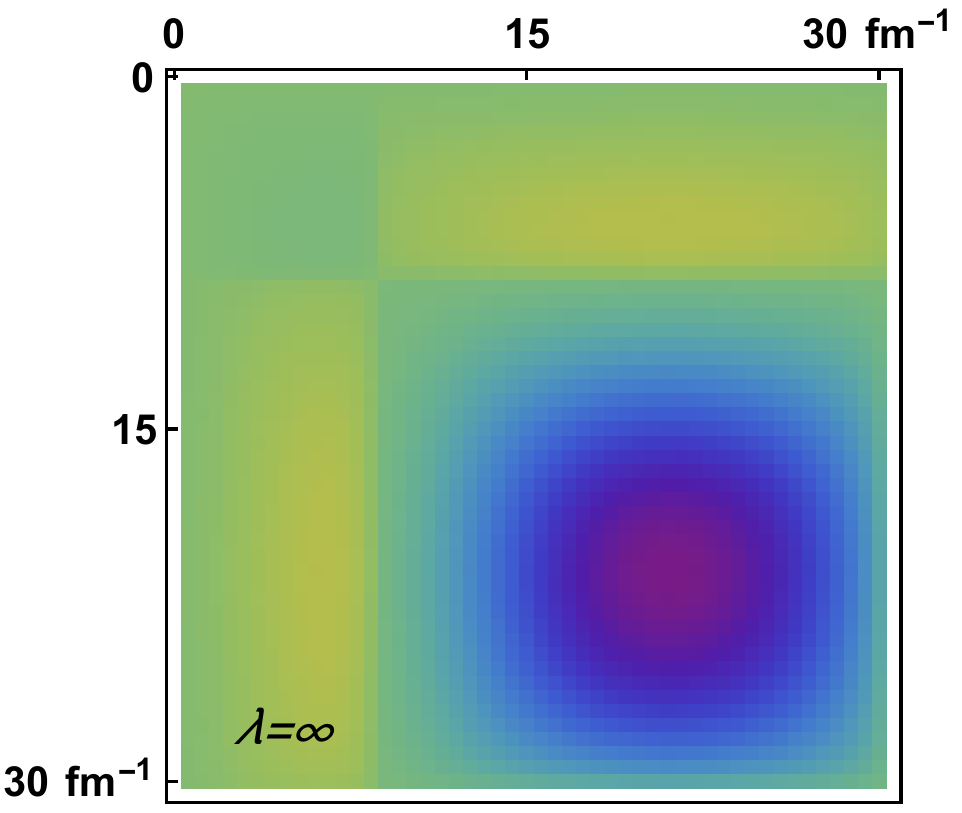}
\includegraphics[scale=0.55]{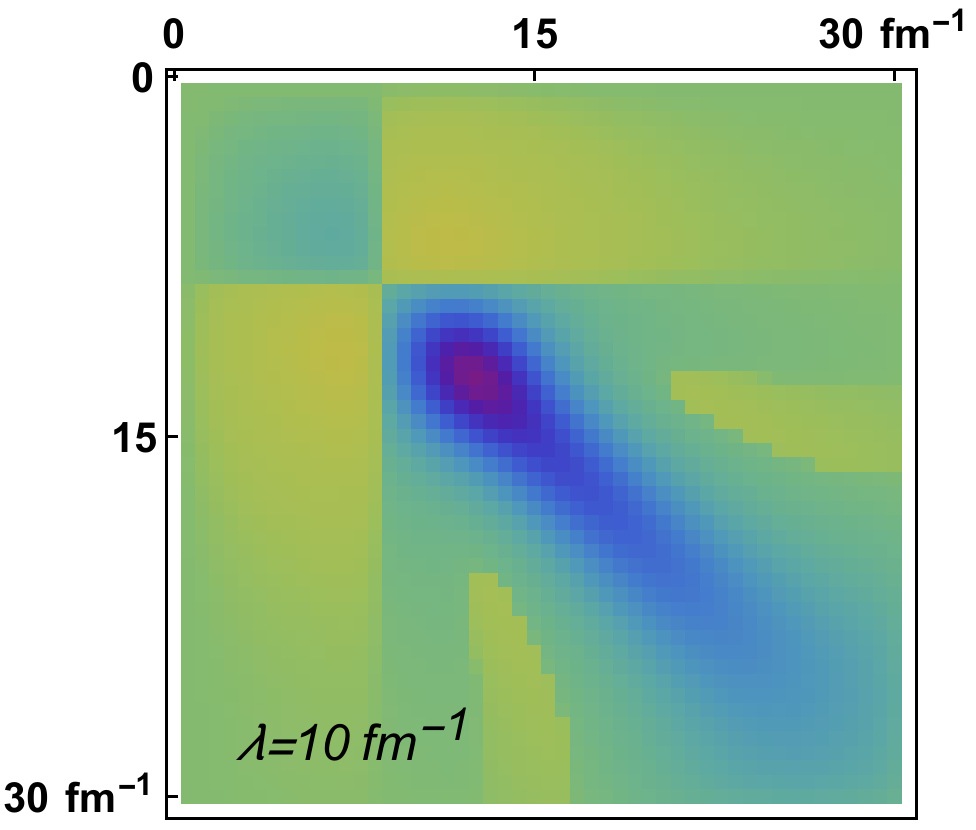}
\includegraphics[scale=0.55]{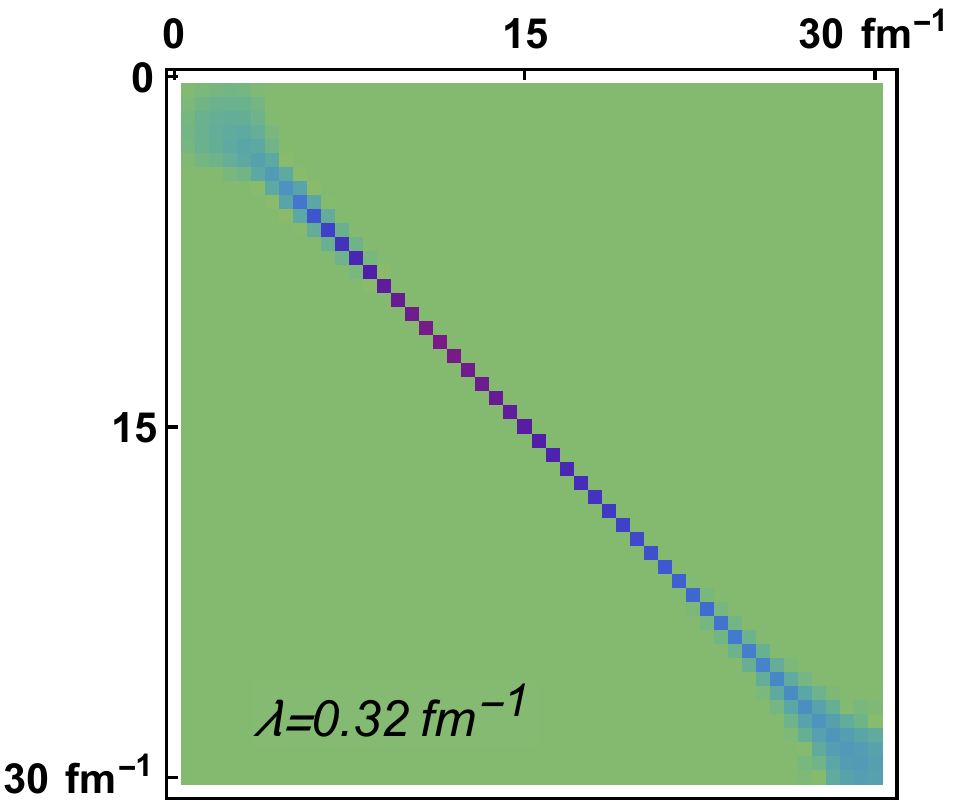}
\\
\includegraphics[scale=0.55]{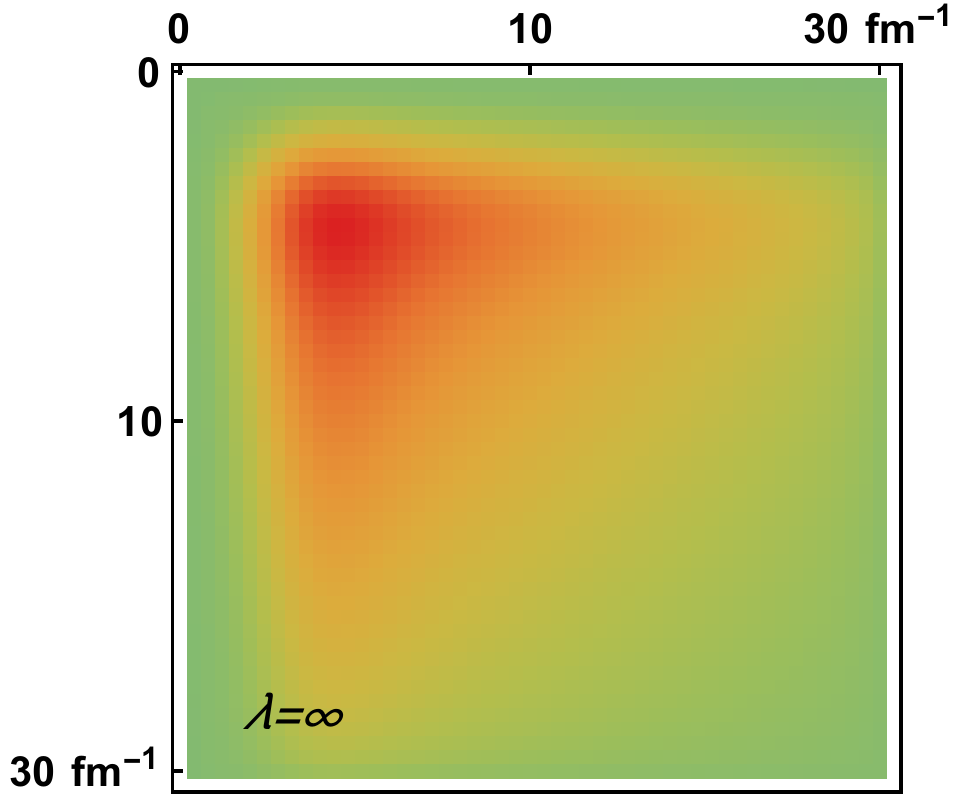}
\includegraphics[scale=0.55]{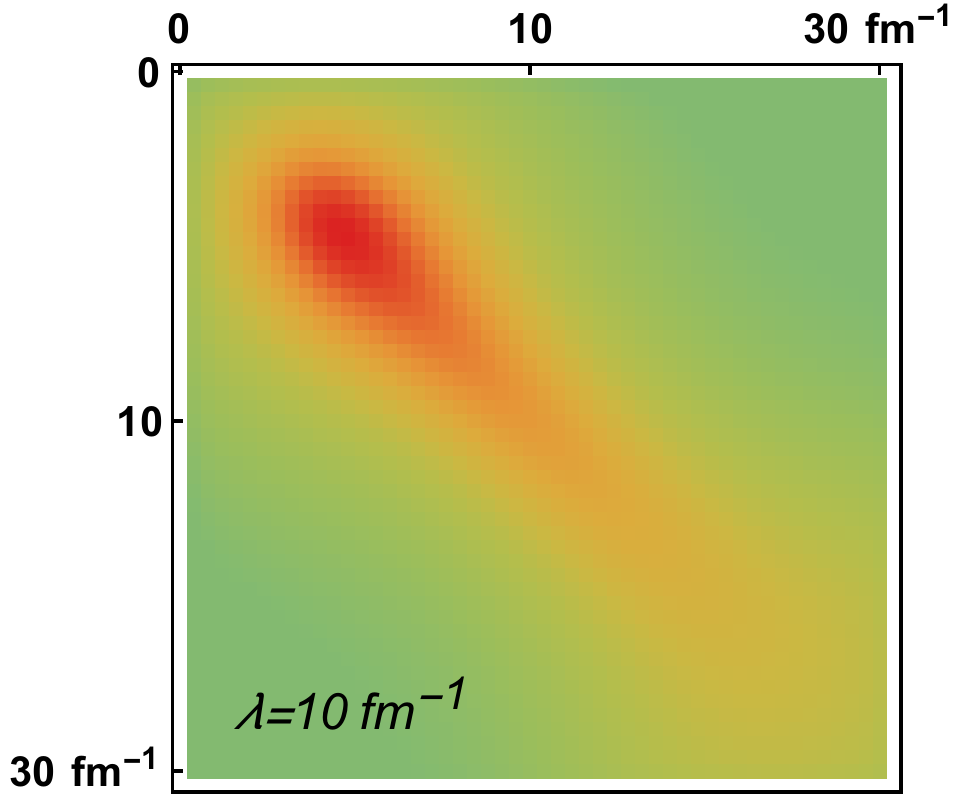}
\includegraphics[scale=0.55]{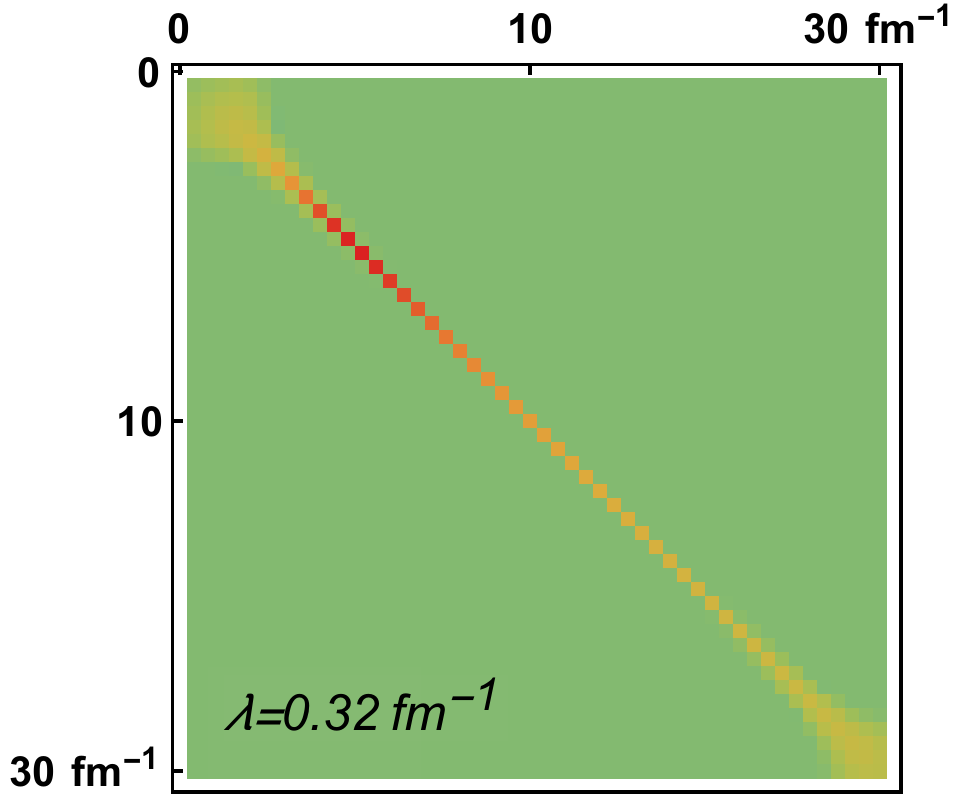}
\caption{Colour online: Evolution of Hamiltonian matrices for three different $\lambda$'s. 
The first, second, and third lines correspond to the $S_0$, $P_1$, $S_2$ waves, respectively. The green colour represents values around zero, while the blue and red tones represent negative and positive values, respectively.  }\label{Fig:evolution}
\end{figure*}

Although the starting Hamiltonian was chosen to be separable, one main effect is that after SRG evolution the potential becomes no longer separable. As mentioned the evolved Hamiltonian preserves the phase-shifts and the SRG evolution provides an explicit example of the 
lack of uniqueness in the determinations of a potential from scattering data. In figure~\ref{Fig:evolution} we present the evolution of states, starting by the initial Hamiltonian ($s=0$), following by $s=0.01$ fm$^{-1}$ and finally for $s=10$ fm$^{-1}$. The first line corresponds to the $S_0$ wave, the second line to the $P_1$ wave, and the third one to the $S_2$ wave. 
In the central column we can appreciate a wide diagonal band. In fact, as advertised, the width of the band is about the value of $\lambda$. Thus, the third column shows a very short band, which is in fact smaller than the difference of values of consecutive matrix elements. 

\section{Conclusion and outlook}

The SRG method, which has been traditionally applied to nuclear interactions and in particular for the nucleon-nucleon potential, 
has been used here for the first time in the description of pion-pion scattering. We have considered the Kadyshevsky equation, that allows for a Hamiltonian interpretation and hence allows for a direct implementation of the SRG equations.  
We have evolved the Hamiltonian of the system using the SRG equation with the Wilson generator corresponding to the kinetic energy. This generator transforms the initial Hamiltonian into a band-diagonal one, in which the matrix elements outside such a band, are negligible. After evolving sufficiently the initial Hamiltonian, one obtains an nearly-diagonal matrix whose diagonal coincide approximately with the eigenvalues of the operator. 
Thus, instead of diagonalizing the Matrix through a finite number of transformations, we have transformed the matris in a continuous way, through a number of infinitesimal transformations in the renormalization-group parameter $s$.
The present result illustrates very simply the nature of SRG methods in hadronic physics. As already discussed, the $\pi\pi$ scattering phase-shifts can be well described by an interaction in momentum space 
with long momentum tails, which fits well but seems unnatural if we consider that the pion is treated as an elementary object. The proper way to address the relevant scales in the problem is by renormalization 
group methods where the energies not relevant to the problem are explicitly integrated out. In the SRG approach this can be simply achieved by considering a block-diagonal generator~\cite{Arriola:2013era,Arriola:2014fqa}. These  
and further issues will be discussed elsewhere~\cite{inpreparation}

\begin{acknowledgments}
We thank Varese Salvador Timoteo for discussions. 
This work has been supported in part by the European Commission under the Marie Sk\l odowska-Curie Action Co-fund 2016 EU project 754446 -- Athenea3i and by the  Spanish  MINECO’s  Juan  de  la  Cierva-Incorporaci\'on programme, Grant Agreement No. IJCI-2017-31531,  FIS2017-8503-C2-1-P and Junta de Andaluc\'ia 
(grant FQM225).
\end{acknowledgments}

\bibliography{aipsamp}

\end{document}